\documentclass[graybox,vecphys]{svmult}

\usepackage{helvet}         % selects Helvetica as sans-serif font
\usepackage{courier}        % selects Courier as typewriter font
\usepackage{type1cm}        % activate if the above 3 fonts are not available on your system
\usepackage{makeidx}         % allows index generation
\usepackage{graphicx}        % standard LaTeX graphics tool when including figure files
\usepackage{multicol}        % used for the two-column index
\usepackage[bottom]{footmisc}% places footnotes at page bottom

\usepackage{color}
\usepackage{amsmath,amssymb}
\usepackage{cite}
\usepackage[dvipdfm,%
colorlinks=true,%
linkcolor=magenta,%
citecolor=blue]{hyperref}

\AtBeginDvi{} %fonts embed

\makeindex             % used for the subject index 
\newcommand{\De}[1]{\mathnormal{\Delta}_{\vect{#1}}}
\newcommand{\En}[1]{E_{\vect{#1}}}
\newcommand{\Ded}[1]{\mathnormal{\Delta}_{\vect{#1}{'}}}
\newcommand{\End}[1]{E_{\vect{#1}{'}}}

\newcommand{\X}[2]{\xi_{\mathrm{#1},\vect{#2}}}
\newcommand{\Xt}[2]{\tilde{\xi}_{\mathrm{#1},\vect{#2}}}
\newcommand{\W}[2]{\epsilon_{\mathrm{#1},\vect{#2}}}
\newcommand{\Wt}[2]{\tilde\epsilon_{\mathrm{#1},\vect{#2}}}
\newcommand{\EX}[1]{\langle #1 \rangle}
\newcommand{\Tr}[1]{\mathrm{Tr}[#1]}
\newcommand{\G}[3]{G^{\mathrm{#1}}_{\mathrm{#2},{\vect{#3}}}}
\newcommand{\vect}[1]{\textbf{\itshape{#1}}}

\def\a0{a_{0}}
\def\pk{p_{\vect{k}}}
\def\nek{n_{\mathrm{e},\vect{k}}}
\def\nhk{n_{\mathrm{h},\vect{k}}}

\def\oa{\hat{a}}
\def\oad{\hat{a}^{\dagger}}
\def\ope{\hat{e}}
\def\oed{\hat{e}^{\dagger}}
\def\oh{\hat{h}}
\def\ohd{\hat{h}^{\dagger}}

\def\ii{\mathrm{i}}
\def\dd{\mathrm{d}}
\begin{document}
\title*{Equilibrium to nonequilibrium condensation in driven-dissipative semiconductor systems}
\titlerunning{Equilibrium to nonequilibrium condensation in driven-dissipative systems}
\author{Makoto Yamaguchi and Tetsuo Ogawa}
\institute{Makoto Yamaguchi \at Department of Physics, Osaka University, 1-1 Machikaneyama, Toyonaka, Osaka 560-0043, Japan \email{yamaguchi@acty.phys.sci.osaka-u.ac.jp}
\and Tetsuo Ogawa \at Department of Physics, Osaka University, 1-1 Machikaneyama, Toyonaka, Osaka 560-0043, Japan \email{ogawa@acty.phys.sci.osaka-u.ac.jp}}

\maketitle

%%------------------------------------------------------------------------
%% Abstract
%%------------------------------------------------------------------------
\abstract{Semiconductor microcavity systems strongly coupled to quantum wells are now receiving a great deal of attention because of their ability to efficiently generate coherent light by the Bose-Einstein condensation (BEC) of an exciton-polariton gas.
Since the exciton polaritons are composite quasi-bosonic particles, many fundamental features arise from their original constituents, i.e., electrons, holes and photons.
As a result, not only equilibrium phases typified by the BEC but also nonequilibrium lasing phases can be achieved.
In this contribution, we describe a framework which can treat such equilibrium and nonequilibrium phases in a unified way.}

%%------------------------------------------------------------------------
%% Introduction
%%------------------------------------------------------------------------
\section{Introduction}\label{sec:intro}
\begin{figure}[tb]%--------- figure -----------
\sidecaption[t]
\includegraphics[width=0.50\linewidth]{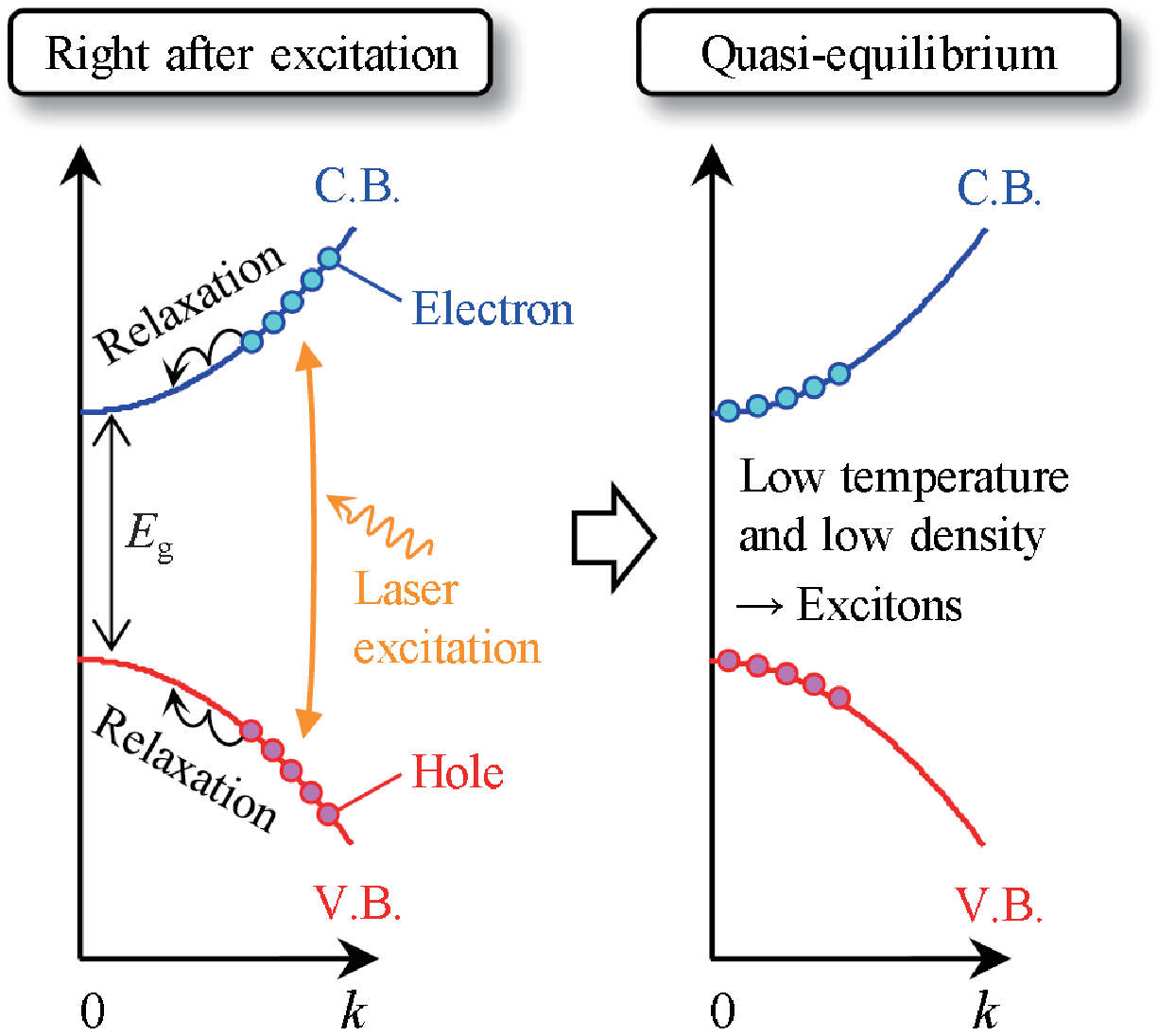}
\caption{
Excitation and thermalization process in a semiconductor.
Electrons and holes generated by laser excitation subsequently undergo immediate intraband relaxations and redistributions in the conduction band (C.B.) and valence band (V.B.) to effectively reach an equilibrium state.
Coulomb-bound e-h pairs (excitons) are formd when the equilibrium state is at sufficiently low temperature and low carrier density.
}
\label{RelaxationRedistribution}
\end{figure}

In a semiconductor system, it is known that electron-hole (e-h) bound pairs can be formed by their Coulomb attraction when the conduction and valence band effectively reach an equilibrium state after the carriers are generated e.g. by laser excitation (Fig.~\ref{RelaxationRedistribution}).
An exciton polariton is a quasi-bosonic particle composed of such a Coulomb-bound e-h pair (exciton) and a photon~\cite{Weisbuch92,Bloch98}, the behaviors of which have attracted much attention due to their potential apprications through the Bose-Einstein condensation (BEC)~\cite{Imamoglu96,Deng02,Kasprzak06}, i.e. a macroscopic occupation of a single exciton-polariton state by a thermodynamic phase transition.

\begin{figure}[tb]%--------- figure -----------
\sidecaption
\includegraphics[width=0.50\linewidth]{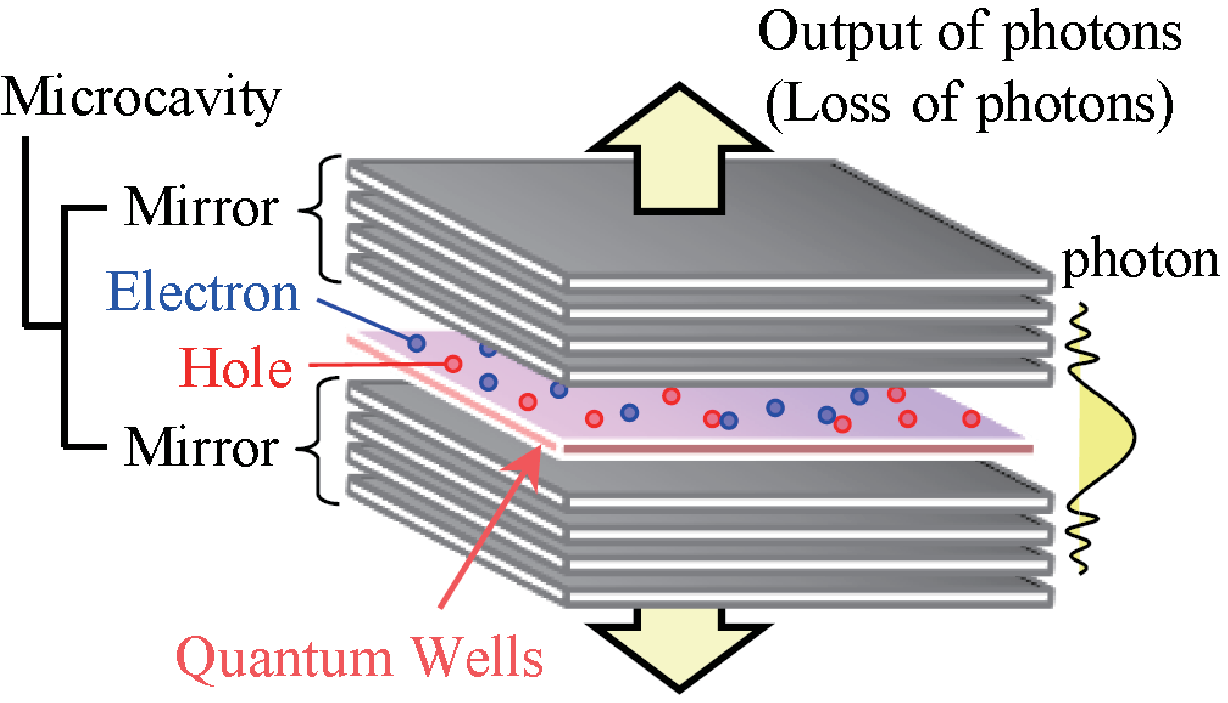}
\caption{
Schematic illustration of a typical exciton-polariton system.
Exciton-polaritons are formed by the electrons and holes in the QWs and the photons confined between the two mirrors (microcavity).
}
\label{System}
\end{figure}

\begin{figure}[tb]%--------- figure -----------
\begin{center}
\includegraphics[width=0.90\linewidth]{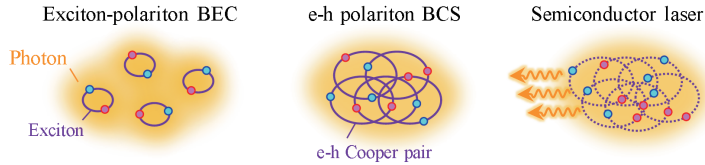}
\end{center}
\vspace{-0.3cm}
\caption{
Schematic illustration of several ordered phases involved in the exciton-polariton system.
}
\label{BEC-BCS-LASER}
\end{figure}

A typical exciton-polariton system is shown in Fig.~\ref{System}.
The system basically consists of semiconductor quantum wells (QWs) and a microcavity, the same structure as a vertical cavity surface emitting laser (VCSEL).
In this context, a conventional lasing phase\footnote{In this contribution, the terms `lasing' and `laser' are used only when the condensation is inherently governed by non-equilibrium kinetics, according to~\cite{Bajoni08,Deng03}.
In other words, thermodynamic variables of the system, such as temperatures, cannot be defined for lasing phases.
However, we note that these terms are occasionally used even for a condensation dominated by the thermodynamics of the electron-hole-photon system~\cite{Kasprzak08} if the interest is in fabricating a device.
}
is involved in this system as well as the exciton-polariton BEC~\cite{Snoke12}.
At high densities, moreover, the Bardeen-Cooper-Schrieffer (BCS) -like ordered phase can potentially be caused where electrons and holes form the ``Cooper pairs''~\cite{Keldysh65,Littlewood04}, as is discussed in the BCS-BEC crossover in cold atom systems with Feshbach resonances~\cite{Ohashi02,Ohashi03}.
These ordered phases are schematically shown in Fig.~\ref{BEC-BCS-LASER}.

However, the BEC and BCS phases are in equilibrium, the situation of which is quite different from the semiconducotor laser in nonequilibrium.
As a result, approaches for describing the BEC and BCS phases based on equilibrium statistical mechanics, e.g. the BCS theory~\cite{Kamide10,Byrnes10}, are not applicable to the semiconductor laser because any nonequilibrium effects cannot be taken into account, such as pumping and loss.
Conversely, past theories for describing the lasing operation, e.g. the Maxwell-Semiconductor-Bloch equations (MSBEs)~\cite{Chow02,Kamide11}, cannot recover such equilibrium statistical approaches.\footnote{We note that theories for describing dynamics of equilibrium phases, e.g. the Gross-Pitaevskii equation, can asymptotically be derived from Maxwell-Bloch equations~\cite{Berloff13} even though these theories still do not recover equilibrium statistical approaches.}
The difficulty shown here has been one of problems to understand the underlying physics in exciton-polariton systems.

In such a situation, we have recently proposed a framework which can treat the phases of the BEC, BCS and laser in a unified way~\cite{Yamaguchi12,Yamaguchi13}.
This framework is an extention of a nonequilibrium Green's function approach developed in Ref.~\cite{Szymanska06,Szymanska07,Keeling10} in which excitons are simply modeled by localized noninteracting two-level systems without internal e-h structures.
Our formalism results in the BCS theory when the system can be regarded as in equilibrium, while it recovers the MSBE when nonequilibrium features become important.
The internal e-h structures as well as the Coulomb interactions can also be taken into account within the mean-field approximation.
In this contribution, we would like to give an introduction to such a ``BEC-BCS-LASER crossover theory''.

%%------------------------------------------------------------------------
%% BCS theory and MSBE for exciton-polariton systems
%%------------------------------------------------------------------------
\section{BCS theory and MSBE for exciton-polariton systems}\label{sec:BCS and MSBE}
In exciton-polariton systems, the equilibrium phases (the BEC and BCS phases) can be described by the BCS theory while the nonequilibrium phase (the lasing phase) can be described by the MSBE.
In this Section, we give an overview of the BCS theory and the MSBE to highlight their similarities and differences.
For simplicity, we set $\hbar=k_{\mathrm{B}}=1$ in the followings.

%---------------------------------
% Model
%---------------------------------
\subsection{Model}\label{subsec:Model}
We first describe the Hamiltonian for the exciton-polariton system where electrons and holes in the QWs and photons in the microcavity are taken into account.
The system Hamiltonian $\hat{H}_{\mathrm{S}}$ is then given by $\hat{H}_{\mathrm{S}} = \hat{H}_{0} + \hat{H}_{\mathrm{Coul}} + \hat{H}_{\mathrm{dip}}$.
Here, $\hat{H}_{\mathrm{0}}$ is the Hamiltonian for free particles without interactions and written as
\begin{equation}
\hat{H}_{\mathrm{0}} = \sum_{\vect{k}} \left( \W{e}{k}\oed_{\vect{k}}\ope_{\vect{k}} + \W{h}{k}\ohd_{\vect{k}}\oh_{\vect{k}} + \W{ph}{k}\oad_{\vect{k}}\oa_{\vect{k}} \right),
\label{eq:HS}
\end{equation}
where $\ope_{\vect{k}}$, $\oh_{\vect{k}}$, and $\oa_{\vect{k}}$ are annihilation operators for electrons, holes, and photons with in-plane wave number $\vect{k}$, respectively.
$\W{e(h)}{k} = k^2/2m_{\mathrm{e(h)}} + E_{\mathrm{g}}/2$ is the energy dispersion of electrons (holes) with an effective mass $m_{\mathrm{e(h)}}$, while $\W{ph}{k} = k^2/2m_{\mathrm{cav}} + E_{\mathrm{cav}}$ is that of photons with an effective mass $m_{\mathrm{cav}}$.
$E_{\mathrm{g}}$ is the bandgap and $E_{\mathrm{cav}}$ is the energy of the cavity mode for $\vect{k}=0$~\cite{Deng10Review}.

In contrast, $\hat{H}_{\mathrm{Coul}}$ and $\hat{H}_{\mathrm{dip}}$ denote the Coulomb interaction and the light-matter interaction within the dipole approximation, respectively written as
\begin{eqnarray}
\hat{H}_{\mathrm{Coul}} &=& \frac{1}{2}\sum_{\vect{k},\vect{k}',\vect{q}}U'_{\vect{q}}
	\left( \oed_{\vect{k}+\vect{q}}\oed_{\vect{k}'-\vect{q}}\ope_{\vect{k}'}\ope_{\vect{k}} + ( \ope \leftrightarrow \oh ) 
	-2\oed_{\vect{k}+\vect{q}}\ohd_{\vect{k}'-\vect{q}}\oh_{\vect{k}'}\ope_{\vect{k}} \right),
\label{eq:HCoul}\\
\hat{H}_{\mathrm{dip}} &=& -\sum_{\vect{k},\vect{q}} \left( g\oad_{\vect{q}}\oh_{-\vect{k}}\ope_{\vect{k}+\vect{q}} + g^{*}\oa_{\vect{q}}\oed_{\vect{k}+\vect{q}}\ohd_{-\vect{k}} \right),
\label{eq:Hdip}
\end{eqnarray}
where $U'_{\vect{q}}=U'_{-\vect{q}}$ and $U'_{\vect{q}=0} \equiv 0$.
Note that $[\hat{H}_{\mathrm{S}},\hat{N}_{\mathrm{S}}] = 0$ is satisfied when an excitation number of the system $\hat{N}_{\mathrm{S}}$ is defined as $\hat{N}_{\mathrm{S}} \equiv \sum_{\vect{k}} [\oed_{\vect{k}}\ope_{\vect{k}}/2 + \ohd_{\vect{k}}\oh_{\vect{k}}/2 + \oad_{\vect{k}}\oa_{\vect{k}}]$.
For later convenience, therefore, we redefine $\hat{H}_{\mathrm{S}}-\mu\hat{N}_{\mathrm{S}}$ as $\hat{H}_{\mathrm{S}}$.
This means that a grand canonical ensemble is assumed with a chemical potential $\mu$ if we are interested in equilibrium phases.
In contrast, for time-dependent problems, this means that dynamics of physical quantities is captured on a rotating frame with the frequency $\mu$.
Thus, $\mu$ is a given parameter identical to the chemical potential for the BEC and BCS phases (Subsection~\ref{subsec:BCS}), whereas it becomes a unknown variable equivalent to the lasing frequency for the semiconductor laser in a steady state (Subsection~\ref{subsec:MSBE}).

%---------------------------------
% Mean-Field approximation
%---------------------------------
\subsection{Mean-field approximation}\label{subsec:MF}
The Hamiltonians shown in Subsection~\ref{subsec:Model} give a starting point for theories of the exciton-polariton system.
However, in practice, it is difficult to exactly treat $\hat{H}_{\mathrm{Coul}}$ and $\hat{H}_{\mathrm{dip}}$ because these Hamiltonians cause many-body problems.
In this Subsection, therefore, we discuss the mean-field (MF) approximation in order to reduce the problems to single-particle problems.

In general, the MF approximation is performed by writing a specific operator $\hat{O}$ as $\hat{O} = \EX{\hat{O}} + \mathnormal{\delta}\hat{O}$ and by neglecting quadratic terms with respect to $\mathnormal{\delta}\hat{O}$ in the Hamiltonians.\footnote{We note, however, that physical guesses are required for the determination of what operator(s) should be chosen as $\hat{O}$, e.g. from experiments.}
Here, $\EX{\hat{O}} \equiv \mathrm{Tr}[\hat{O}\hat\rho]$ denotes the expectation value for the density operator $\hat\rho$ and the operator $\mathnormal{\delta}\hat{O}$ corresponds to a fluctuation around the expectation value.
In our case, the interaction Hamiltonians of $\hat{H}_{\mathrm{Coul}}$ and $\hat{H}_{\mathrm{dip}}$ can easily be reduced to a single-particle problem by employing $\hat{O} \in \{ \oa_{\vect{k}}, \oh_{-\vect{k}}\ope_{\vect{k}'}, \oed_{\vect{k}}\ope_{\vect{k}'}, \ohd_{\vect{k}}\oh_{\vect{k}'} \}$.
As a result, with definitions of the photon field $\EX{\oa_{\vect{k}}} \equiv \delta_{\vect{k},0}\a0$,
the polarization function $\EX{\oh_{-\vect{k}}\ope_{\vect{k}'}} \equiv \delta_{\vect{k},\vect{k}'}\pk$,
and the distribution functions of electrons $\EX{\oed_{\vect{k}}\ope_{\vect{k}'}} \equiv \delta_{\vect{k},\vect{k}'}\nek$
and holes $\EX{\ohd_{\vect{k}}\oh_{\vect{k}'}} \equiv \delta_{\vect{k},\vect{k}'}\nhk$, the mean-field Hamiltonian $\hat{H}_{\mathrm{S}}^{\mathrm{MF}}$ is obtained as
\begin{eqnarray}
\hat{H}_{\mathrm{S}}^{\mathrm{MF}} &=& \sum_{\vect{k}} \left( \Xt{e}{k}\oed_{\vect{k}}\ope_{\vect{k}} + \Xt{h}{k}\ohd_{\vect{k}}\oh_{\vect{k}} - [\De{k}\oed_{\vect{k}}\ohd_{-\vect{k}} + \mathrm{H.c.}] \right) \nonumber\\
&&+\sum_{\vect{k}} \left( \X{ph}{k}\oad_{\vect{k}}\oa_{\vect{k}} - [g\pk\oad_0 + g^{*}\pk^{*}\oa_0] \right).
\label{eq:HMF}
\end{eqnarray}
Here, constants are ignored because the following discussion is not affected.
$\Xt{e(h)}{k}$ and $\X{ph}{k}$ are respectively defined as $\Xt{e(h)}{k} \equiv \Wt{e(h)}{k} - \mu/2$ and $\X{ph}{k} \equiv \W{ph}{k} - \mu$, where $\Wt{e(h)}{k} \equiv \W{e(h)}{k} - \sum_{\vect{k}'} U'_{\vect{k}-\vect{k}'} n_{\mathrm{e(h)},\vect{k}'}$ denotes the energy dispersion of electrons (holes) renormalized by the repulsive electron-electron (hole-hole) Coulomb interaction, the first (second) term in Eq.~(\ref{eq:HCoul}).
The well-known bandgap renormalization (BGR) in semiconductor physics is included in $\Wt{e(h)}{k}$.
In contrast, $\De{k} \equiv g^{*}a_0 + \sum_{\vect{k}'}U'_{\vect{k}-\vect{k}'}p_{\vect{k}'}$ results from the attractive electron-hole Coulomb interaction, the third term in Eq.~(\ref{eq:HCoul}), and is called the generalized Rabi frequency~\cite{HaugKoch}.
$\De{k}$ has a role in forming e-h pairs as can be seen in Eq.~(\ref{eq:HMF}).

The mean-field Hamiltonian is thus obtained.
However, note that the expectation values of $\EX{\hat{O}}$ (i.e. $\a0$, $\pk$, $\nek$, and $\nhk$) are included in $\hat{H}_{\mathrm{S}}^{\mathrm{MF}}$.
For self-consistensy, therefore, the following relation should be satisfied :
\begin{equation}
\EX{\hat{O}} = \Tr{\hat{O}\hat\rho^{\mathrm{MF}}(\EX{\hat{O}})}.
\label{eq:SCE}
\end{equation}
Here, $\hat\rho^{\mathrm{MF}}$ is the density operator determiend by using $\hat{H}_{\mathrm{S}}^{\mathrm{MF}}$.
The BCS theory and the MSBE shown below are obtained from this self-consistent equation.

%---------------------------------
% BCS theory for exciton-polariton systems
%---------------------------------
\subsection{BCS theory for exciton-polariton condensation}\label{subsec:BCS}
First, we assume that the exciton-polariton system is in equilibrium. 
According to the equilibrium statistical mechanics, the density operator $\hat\rho^{\mathrm{MF}}$ at temperature $T$ can be described as
\begin{eqnarray}
\hat\rho^{\mathrm{MF}}=\hat\rho_{\mathrm{eq}}^{\mathrm{MF}}\equiv\frac{1}{Z}\exp(-\beta\hat{H}_{\mathrm{S}}^{\mathrm{MF}}),
\label{eq:equilibrium}
\end{eqnarray}
where $Z \equiv \mathrm{Tr}[\exp(-\beta\hat{H}_{\mathrm{S}}^{\mathrm{MF}})]$ and $\beta \equiv 1/T$.
In this case, $\mu$ is a given parameter equivalent to the chemical potential, as mentioned above.
With $\W{e}{k} = \W{h}{k}$ for simplicity, the self-consistent equations obtained from Eqs.~(\ref{eq:HMF})-(\ref{eq:equilibrium}) are
\begin{eqnarray}
\a0=\sum_{\vect{k}'}\frac{g}{\xi_{\mathrm{ph},0}}p_{\vect{k}'},\ \pk=\frac{\De{k}}{2\En{k}}\tanh \left( \frac{\beta \En{k}}{2} \right),
\label{eq:ap}\\
\nek=\nhk=\frac{1}{2} \left\{ 1-\frac{\Xt{eh}{k}^{+}}{\En{k}}\tanh \left( \frac{\beta \En{k}}{2} \right)  \right\},
\label{eq:number}
\end{eqnarray}
where $\Xt{eh}{k}^{\pm} \equiv (\Xt{e}{k} \pm \Xt{h}{k})/2 $ and $\En{k} \equiv [(\Xt{eh}{k}^{+})^2 + |\De{k}|^2]^{1/2}$.
In the derivation, Bogoliubov transformations of $\ope_{\vect{k}}$ and $\oh_{\vect{k}}$ can be applied to the first line in Eq.~(\ref{eq:HMF}) for diagonalization, while a displacement of $\oa_{0}$ to the second line, because the Hilbert space of the first (second) line of Eq.~(\ref{eq:HMF}) is spanned only by the electron and hole (photon) degrees of freedom.

The gap equation, which is formally equivalent to the BCS theory for superconductors, can then be obtained by substituting Eq.~(\ref{eq:ap}) into the definition of $\De{k}$:
\begin{equation}
\displaystyle \De{k} = \sum_{\vect{k}'}U_{\vect{k}',\vect{k}}^{\mathrm{eff}}\frac{\Ded{k}}{2\End{k}} \tanh \left( \frac{\beta \End{k}}{2} \right).
\label{eq:gap}
\end{equation}
In this context, $\De{k}$ is an order parameter in the exciton-polariton system as well as in the superconducting system.
$U_{\vect{k}',\vect{k}}^{\mathrm{eff}} \equiv |g|^2/\xi_{\mathrm{ph},0} + U'_{\vect{k}'-\vect{k}}$ represents an effective attractive e-h interaction, from which one can find that photon-mediated process also contributes the attractive interaction.
Notice that Eqs.~(\ref{eq:number}) and (\ref{eq:gap}) are simultaneous equations with the unknown variables $\nek (=\nhk)$ and $\De{k}$.
Especially for $T = 0$, this treatment is known to cover the equilibrium phases from the BEC to the BCS states~\cite{Kamide10,Byrnes10,Comte82}.

\begin{figure}[b]%--------- figure -----------
\sidecaption
\includegraphics[width=0.50\linewidth]{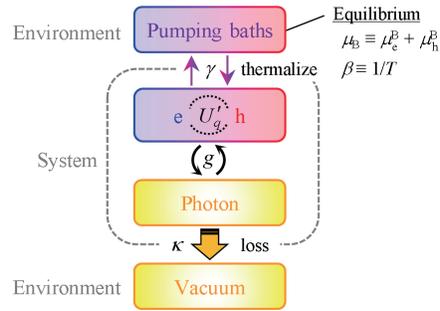}
\caption{
A schematic picture of the model including environments.
The e-h system is excited and thermalized by the pumping baths by exchanging carriers.
Photons in the system are lost into the vacuum. 
}
\label{Model}
\end{figure}

%---------------------------------
% MSBE for semiconductor lasers
%---------------------------------
\subsection{MSBE for semiconductor lasers}\label{subsec:MSBE}
\begin{figure}[ht]%--------- figure -----------
\sidecaption
\includegraphics[width=0.50\linewidth]{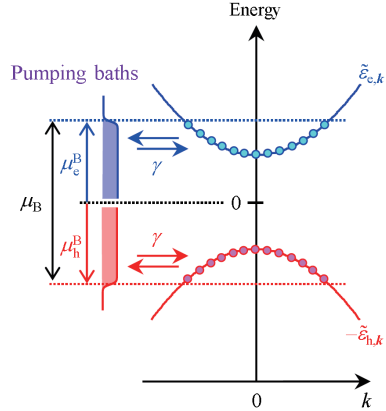}
\caption{
Energy dispersions of electrons and holes.
The distribution functions $n_{\mathrm{e},\vect{k}}$ and $n_{\mathrm{h},\vect{k}}$ are driven to approach the Fermi distributions by the respective pumping baths.
}
\label{Energy}
\end{figure}

Next, a treatment based on the MSBE is explained for the discussion of the semiconductor laser, which is characterized by nonequilibrium.
In contrast to the BCS theory, therefore, the effects of environments (Fig.~\ref{Model}) cannot be neglected for lasing; the excitation and thermalization of the e-h system and the loss of photons from the microcavity.
For this reason, the dynamics of the total density operator $\hat\rho^{\mathrm{MF}}$ is discussed by writing the total mean-field Hamiltonian $\hat{H}^{\mathrm{MF}} \equiv \hat{H}_{\mathrm{S}}^{\mathrm{MF}}+\hat{E}$ with the couplings to the environments $\hat{E}$.
Since $\ii\partial_t \hat\rho^{\mathrm{MF}} = [\hat{H}^{\mathrm{MF}},\hat\rho^{\mathrm{MF}}]$ in the Schr\"{o}dinger picture, a time drivative of Eq.~(\ref{eq:SCE}) yields
\begin{equation}
\ii\partial_t\EX{\hat{O}} = \Tr{[\hat{O},\hat{H}_{\mathrm{S}}^{\mathrm{MF}}]\hat\rho^{\mathrm{MF}}}+\Tr{[\hat{O},\hat{E}]\hat\rho^{\mathrm{MF}}},
\label{eq:SCE2}
\end{equation}
where $\Tr{\hat{A}\hat{B}}=\Tr{\hat{B}\hat{A}}$ is used.
The MSBE is then obtained when the first term is derived from Eq.~(\ref{eq:HMF}) and the second term is replaced by phenomenological relaxation terms:
\begin{eqnarray}
\partial_t\a0&=&-\ii \xi_{\mathrm{ph},0} \a0 + \ii g \textstyle{\sum_{\vect{k}}} \pk - \kappa \a0,
\label{eq:MSBE_a0} \\
\partial_t\pk&=&-2\ii\Xt{eh}{k}^{+}\pk-\ii\De{k} N_{\vect{k}}-2\gamma (\pk - \pk^{0}),
\label{eq:MSBE_pk} \\
\partial_t n_{\mathrm{e(h)},\vect{k}}&=&-2\mathrm{Im}[\De{k} p_{\vect{k}}^{*}]-2\gamma(n_{\mathrm{e(h)},\vect{k}}-n_{\mathrm{e(h)},\vect{k}}^{0}),
\label{eq:MSBE_nehk}
\end{eqnarray}
where the last term in each equation is the relaxation term and $N_{\vect{k}} \equiv \nek + \nhk -1$ denotes the degree of the population inversion.\footnote{Here, $-1 \le N_{\vect{k}} \le +1$ because $0 \le n_{\mathrm{e(h)},\vect{k}} \le 1$. Population inversion is formed in $\vect{k}$-resions with $N_{\vect{k}}>0$.}
$\pk^{0}$ and $n_{\mathrm{e(h)},\vect{k}}^{0}$ are defined as
\begin{equation}
\pk^{0} \equiv 0, \hspace{5pt} n_{\mathrm{e(h)},\vect{k}}^{0} \equiv f_{\mathrm{e(h)},\vect{k}} 
\label{eq:RTA}
\end{equation}
where $f_{\mathrm{e(h)},\vect{k}} \equiv [1 + \exp \{ \beta ( \Wt{e(h)}{k}-\mu^{\mathrm{B}}_{\mathrm{e(h)}}) \} ]^{-1}$ is the Fermi distribution with the chemical potential $\mu^{\mathrm{B}}_{\mathrm{e(h)}}$ of the electron (hole) pumping bath.
The phenomenological approximation shown here is called the relaxation approximation~\cite{Henneberger92}.
Each relaxation term suggests that the photon field $a_{0}$ decays with a rate of $\kappa$, the distribution function $n_{\mathrm{e(h)},\vect{k}}$ is driven to approach the Fermi distribution $f_{\mathrm{e(h)},\vect{k}}$ (Fig.~\ref{Energy}), i.e. thermalization (Fig.~\ref{RelaxationRedistribution}), and $\pk$ decays due to thermalization-induced dephasing.

Solutions for the laser action can then be obtained by determining the unknown variables $\a0$, $\pk$, $\nek$, $\nhk$, and $\mu$ in Eqs.~(\ref{eq:MSBE_a0})-(\ref{eq:MSBE_nehk}) under a steady-state condition $\partial_t\EX{\hat{O}} = 0$.
Again, we emphasize that $\mu$ is a unknown variable corresponding to the laser frequency in the steady-state MSBE, in contrast to the BCS theory.
This is equivalent to find an appropriate frequency with which the lasing oscillation of $\a0$ and $\pk$ seems to remain stationary on the rotating frame.

%%------------------------------------------------------------------------
%% BEC-BCS-LASER crossover theory
%%------------------------------------------------------------------------
\section{BEC-BCS-LASER crossover theory}\label{sec:BBL}
In the exciton-polariton system, as shown in Section~\ref{sec:BCS and MSBE}, the BCS theory and the MSBE are theoretical frameworks starting from the common Hamiltonians with the same mean-field approximation.
However, the difference is the way of deriving the self-consistent equations.
In the case of the BCS theory, $\hat\rho^{\mathrm{MF}}$ is directly described by $\hat{H}_{\mathrm{S}}^{\mathrm{MF}}$ (Eq.~(\ref{eq:equilibrium})).
In contrast, in the case of the MSBE, Eq.~(\ref{eq:SCE2}) is used to introduce the phenomenological relaxation terms.
We note, however, that any assumption is not used for $\hat\rho^{\mathrm{MF}}$ in Eq.~(\ref{eq:SCE2}), which indicates that the MSBE may incorporate the BCS theory at least in principle. 

In this context, an approach to derive the BCS theory from the MSBE should be discussed briefly.
We first consider a situation where the effects of the environments are completely neglected, which is equivalent to set $\kappa = \gamma = 0$ in the MSBE.
However, in this case, the BCS theory cannnot be derived because there is no term to drive the system into equilibrium in the MSBE.\footnote{For $\kappa = \gamma = 0$, the steady state of the MSBE becomes identical to the BCS theory if the solution of the BCS theory is chosen as an initial condition in the MSBE because $[\hat{\rho}_{\mathrm{eq}}^{\mathrm{MF}},\hat{H}_{\mathrm{S}}^{\mathrm{MF}}]=0$.
However, this is a special case.
}
A natural condition to consider is physically a limit of $\gamma \rightarrow 0^{+}$ after $\kappa \rightarrow 0$ because the system should be thermalized even though the effects of environments are decreased.
Unfortunately, however, the MSBE does not recover the BCS theory even by taking this limit.
The relationship between the BCS theory and the MSBE is thus discontinuous in spite of the similarities of the two frameworks.
Obviously, the phenomenological relaxation approximation causes such a problem.

In regard to this problem, we have recently constructed a unified framework~\cite{Yamaguchi12} by using a nonequilibrium Green's function approach~\cite{Szymanska06,Szymanska07,Keeling10}.
The framework, at first, takes an integral form of simultaneous equations and seems quite different from the MSBE (see also Appendix I\hspace{-.1em}I).
However, by rearranging the equations with particular attention to the problem mentioned above, all of important changes can successfully be incorporated in the relaxation terms in the MSBE~\cite{Yamaguchi13}.
The result simply replaces Eq.~(\ref{eq:RTA}) by 
\begin{align}
&\pk^{0} \equiv \ii\int\frac{\mathrm{d}\nu}{2\pi}\left[ \G{R}{12}{k}(\nu)\{1-f^{\mathrm{B}}_{\mathrm{h}}(-\nu)\} - \G{R*}{21}{k}(\nu) f^{\mathrm{B}}_{\mathrm{e}}(\nu) \right], \nonumber\\
&n_{\mathrm{e(h)},\vect{k}}^{0} \equiv \int\frac{\mathrm{d}\nu}{2\pi}f^{\mathrm{B}}_{\mathrm{e(h)}}(\nu)A_{\mathrm{11(22)}}(\pm\nu;\vect{k}),
\tag{\ref{eq:RTA}$'$} \label{eq:RTA'}
\end{align}
where $f^{\mathrm{B}}_{\mathrm{e(h)}}(\nu) \equiv [\exp \{ \beta(\nu - \mu^{\mathrm{B}}_{\mathrm{e(h)}} + \mu/2) \} + 1]^{-1}$ is the Fermi distribution of the electron (hole) pumping bath.
$\G{R}{\alpha\alpha'}{k}(\nu)$ is called the retarded Green's function and described by elements of a matrix
\begin{eqnarray}
G^{\mathrm{R}}_{\vect{k}}(\nu)=\left(
\begin{array}{cc}
\nu - \tilde{\xi}_{\mathrm{e},\vect{k}} + \ii\gamma & \De{k} \\
\De{k}^{*} & \nu + \tilde{\xi}_{\mathrm{h},\vect{k}} + \ii\gamma \\
\end{array}\right)^{-1}.
\label{eq:GR}
\end{eqnarray}
On the other hand, $A_{11(22)}(\nu;\vect{k})$ is called the single-particle spectral function and defined as 
\begin{equation}
A_{\alpha\alpha'}(\nu;\vect{k}) \equiv \ii (G^{\mathrm{R}}_{\alpha\alpha',\vect{k}}(\nu) - G^{\mathrm{R}*}_{\alpha'\alpha,\vect{k}}(\nu)).
\label{eq:A}
\end{equation}
Here, $A_{11(22)}(\nu;\vect{k})$ means the density of states for electron-like (hole-like) quasi-particles with the energy $\nu$ and wave number $\vect{k}$.

Some readers might feel difficult to understand the formalism because the above definitions are unique to the Green's function approach.
However, all we have to do is the replacement of Eq.~(\ref{eq:RTA}) by Eq.~(\ref{eq:RTA'}).
The unknown variables are still $\a0$, $\pk$, $\nek$, $\nhk$, and $\mu$, that is, the same as the MSBE.
In this sense, the obtained equations are quite simple, which is one of strong points of this formalism.
From the viewpoint of the Green's funciton, it is relatively easy to understand the physical meaning of Eq.~(\ref{eq:RTA'}) due to the clear form; the energy integral of (distribution)$\times$(density of states).\footnote{The retarded Green's funcition is also seen as a kind of density of states.}
We refer to such a formalism as the BEC-BCS-LASER crossover theory.

Now, this formalism enables us to cleary understand the standpoint of the BCS theory.
For this purpose, let us discuss the limit of equilibrium, based on the idea described above.
In the followings, however, $\W{e}{k} = \W{h}{k}$ and a charge neutrality $\mu^{\mathrm{B}}_{\mathrm{e}} = \mu^{\mathrm{B}}_{\mathrm{h}}$ are assumed for simplicity.
First, in the limit of $\kappa \rightarrow 0$, one can prove $\mu = \mu_{\mathrm{B}} (\equiv \mu^{\mathrm{B}}_{\mathrm{e}} + \mu^{\mathrm{B}}_{\mathrm{h}})$.
This is the same as treating $\mu$ as a given parameter, and physically, means that the system reaches in chemical equilibrium with the pumping baths because there is no photon loss.
The BCS theory is then derived after taking the limit of $\gamma \rightarrow 0^{+}$, where the integrals in Eq.~(\ref{eq:RTA'}) can be performed analytically.
In this derivation, $\gamma \neq 0$ is required to be canceled down even though $\gamma$ does not appear in the final expression.
This means that thermalization is essential to recover the equilibrium theory.

Thus, the BCS theory can be derived from the presented theory in the equilibrium limit.
However, in some sense, this situation is physically trivial; the situation is not limited to such a trivial one for the system to be in equilibrium.
Even under a condition where photons are continuously lost, it may be still possible to identify the system as being in equilibrium (quasi-equilibrium) as long as the e-h system is excited and thermalized.
A true advantage of the above-presented framework becomes obvious in such a situation rather than in the trivial one.
In this case, $\mu$ is still equivalent to the chemical potential but $\mu_{\mathrm{B}} > \mu$ because the system is influenced by the photon loss.
As a result, $\mu$ becomes a unknown variable again.
Furthermore, such a quasi-equilibrium condition can easily be obtained from Eqs.~(\ref{eq:RTA'})-(\ref{eq:A}) as\footnote{
Under the condition~(I), $f^{\mathrm{B}}_{\mathrm{e(h)}}(\nu)$ in Eq.~(\ref{eq:RTA'}) can be approximated by the values at $\nu = \pm \En{k}$ because $A_{\alpha\alpha'}(\nu;\vect{k})$ and $G^{\mathrm{R}}_{\alpha\alpha',\vect{k}}(\nu)$ have peaks around $\nu = \pm \En{k}$, as seen in Eqs.~(\ref{eq:A2}) and (\ref{eqa:GR}).}
\begin{center}
(I)~$\min[2\En{k}] \gtrsim \mu_{\mathrm{B}} - \mu + 2\gamma + 2T$.
\end{center}
Here, $\min[2\En{k}]$ is the minimum energy required for breaking e-h bound pairs and $\mu_{\mathrm{B}}-\mu > 0$ suggests that there is continuous particle flow from the pumping baths into the system.\footnote{This means that the system is chemically non-equilibrium with the pumping baths even if the system is in quasi-equilibrium.}
We can then interpret the condition (I); this is a condition that the particle flux, thermalization-induced dephasing (= $2\gamma$), and temperature effect (= $2T$), do not contribute to the dissociations of the e-h pairs.

However, the system can no longer be in quasi-equilibrium when nonequilibrium effect becomes significant.
Let us therefore consider a situation where the MSBE, i.e. the physics of the semiconductor laser, becomes important.
Such a condition can be found from Eqs.~(\ref{eq:RTA'})-(\ref{eq:A}) as 
\begin{center}
(II)~$\mu_{\mathrm{B}} - \mu \gtrsim \min[2\En{k}] + 2\gamma + 2T$,
\end{center}
because $f^{\mathrm{B}}_{\mathrm{e(h)}}(\pm\nu) \simeq f^{\mathrm{B}}_{\mathrm{e(h)}}(\Xt{eh}{k}^{+})$ turns out to be a good approximation in Eq.~(\ref{eq:RTA'}) for $\vect{k}$-resions satisfying
\begin{center}
(II$'$)~$~\mu_{\mathrm{B}} - \mu \gtrsim 2\En{k} + 2\gamma + 2T$.
\end{center}
Note that there are such $\vect{k}$-resions whenever the condition (II) is fulfilled.
As a result, we can obtain $p_{\vect{k}}^0 \cong 0$ and $n_{\mathrm{e(h)},\vect{k}}^0 \cong f_{\mathrm{e(h)},\vect{k}}$ which recovers the MSBE.
However, we stress that the condition (II$'$) depends on the wave number $\vect{k}$; there remain $\vect{k}$-regions still described by the BCS theory.
The MSBE and the BCS theory are, thus, coupled with each other in a strict sense.
In this context, the lasing can be referred to as the BCS-coupled lasing when this viewpoint is emphasized.
At the same time, the physical meaning of $\mu$ changes into the oscillating frequency of the laser action.

%%------------------------------------------------------------------------
%% Second thresholds and gain spectra
%%------------------------------------------------------------------------
\section{Second thresholds, band renormalization, and gain spectra}\label{sec:Results}
\begin{figure}[tbh]%--------- figure -----------
\begin{center}
\includegraphics[width=0.70\linewidth]{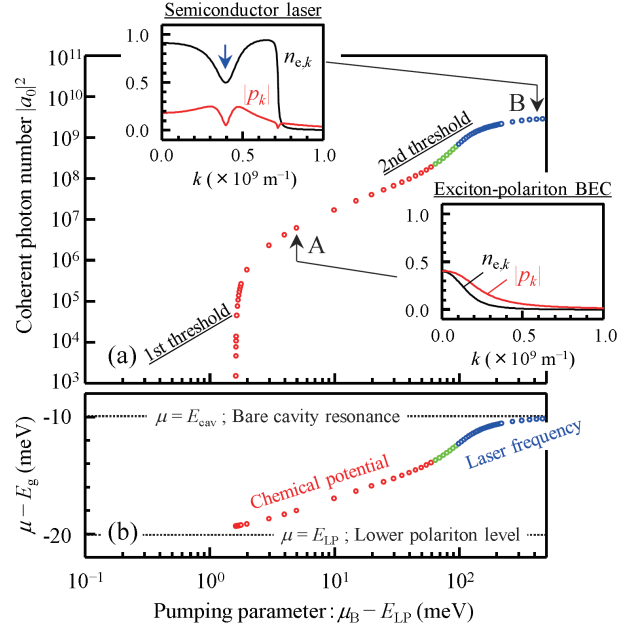}
\end{center}
\vspace{-0.3cm}
\caption{
Numerical results of (a) the coherent photon number in the cavity $|a_{0}|^2$ and (b) the frequency $\mu$ as a function of $\mu_{\mathrm{B}}$.
Plots are colour coded by red and blue when satisfying the quasi-equilibrium condition (I) and the lasing condition (II), respectively.
Green colours are used when neither of them are satisfied.
$\mu$ represents the chemical potential in the quasi-equilibrium regime (red) but the laser frequency in the lasing regime (blue).
Inset: the distribution function of electrons $n_{\mathrm{e},\vect{k}}$ (black) and the polarization $\pk$ (red).
In the lasing regime (B), a characteristic dip can be seen in the distribution (the bule arrow), which is known as one of the signatures of lasing and called the kinetic hole burning.
}
\label{Results}
\end{figure}

Figure~\ref{Results} shows the number of coherent photons in the cavity $|a_{0}|^2$ and the frequancy $\mu$ as a function of $\mu_{\mathrm{B}}$ calculated by our formalism.\footnote{
In the numerical calculations, the $\vect{k}$-dependence of $\De{k}$ is eliminated by using a contact potential $U'_{\vect{q} \neq 0} = U = 2.66 \times 10^{-10}$~eV with cut-off wave number $k_{\mathrm{c}} = 1.36 \times 10^{9}$ $m^{-1}$.
The other parameters are $m_{\mathrm{e}} = m_{\mathrm{h}} = 0.068m_0$ ($m_0$ is the free electron mass), $\mu^{\mathrm{B}}_{\mathrm{e}} = \mu^{\mathrm{B}}_{\mathrm{h}}$, $T = 10 $K, $g = 6.29 \times 10^{-7}$eV, $\gamma = 4$meV, and $\kappa = 100 \mathrm{\mu eV}$.
In this context, our calculations are not quantitative but qualitative even though the parameters are taken as realistic as possible.
In this situation, the exciton level ($\equiv E_{\mathrm{ex}}$) is formed at 10~meV below $E_{\mathrm{g}}$ ($E_{\mathrm{ex}} = E_{\mathrm{g}} - 10$ meV) and the lower polariton level $E_{\mathrm{LP}}$ is created at 20 meV below $E_{\mathrm{g}}$ ($E_{\mathrm{LP}} = E_{\mathrm{g}} - 20$ meV) under the resonant condition $E_{\mathrm{cav}} = E_{\mathrm{ex}}$~\cite{Yamaguchi13}.
}
Plots are colour coded by red (blue) when the quasi-equilibrium condition (I) (the lasing condition (II)) is satisfied, while by green when neither of the conditions is satisfied.
In Fig.~\ref{Results}(a), $|a_{0}|^2$ arises with increasing $\mu_{\mathrm{B}}$, the point of which is called the first threshold. 
In this situation, the system is in quasi-equilibrium regime (red) and $\mu$ is around the lower polariton level\footnote{For $E_{\mathrm{LP}}$ and $E_{\mathrm{ex}}$, see also Appendix I. Excitonic effects are discussed in the low density limit.} $E_{\mathrm{LP}}$ in Fig.~\ref{Results}(b).
The first threshold therefore means that the exciton-polariton BEC is caused because the chemical potential of the system reaches the lowest energy of the exciton polariton, $E_{\mathrm{LP}}$.

With further increase of $\mu_{\mathrm{B}}$, the system changes from the quasi-equilibrium regime (red) into the lasing regime (blue) through a crossover regime (green).
Around the crossover regime in Fig.~\ref{Results}(a), a second threshold can be seen where the number of coherent photons grows rapidly again.
$\mu$ is then blue-shifted from $E_{\mathrm{LP}}$ into the bare cavity level $E_{\mathrm{cav}}$.
Furthermore, the kinetic hole burning can be seen in the distribution function of electrons $\nek$ (the blue arrow in the inset to Fig.~\ref{Results}(a)).
These resuls demonstrate that the exciton-polariton BEC has smoothly changed into the semiconductor laser with the second threshold.

In experiments~\cite{Balili09,Nelsen09,Dang98,Tempel12-1,Tempel12-2,Tsotsis12,Kammann12}, the second threshold and the blue shift has been reported since more than 10 years ago, the mechanism of which has been attributed to a shift into the weak coupling regime due to dissociations of Coulomb-bound e-h pairs (excitons); the lasing phase is then achieved as a result.
However, there is no convincing discussion why such dissociations lead to nonequilibration essential for lasing.

\begin{figure}[htb]%--------- figure -----------
\begin{center}
\includegraphics[width=0.80\linewidth]{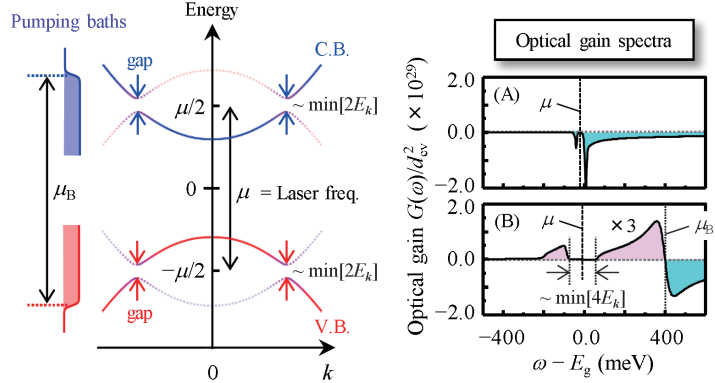}
\end{center}
\vspace{-0.3cm}
\caption{Left; A renormalized band structure in a lasing phase (the point B in Fig.~\ref{Results}).
The gaps are opened around $\pm\mu/2$ with the magnitude of $\min[2\En{k}]$.
Right; Optical gain spectra for the exciton-polariton BEC (A) and for the lasing phase (B).
Panels (A) and (B) correspond to the point A and B in Fig.~\ref{Results}, respectively.
Aqua (pink) represents the gain (absorption).
Panels (A) and (B) are reproduced from \cite{Yamaguchi13}.
}
\label{Gain}
\end{figure}

According to our formalism, this empirical picture can be investigated and shown to be incorrect.
This is because, even in the lasing regime, there are gaps around $\pm\mu/2$ in the renormalized band structure as shown in the left of Fig.~\ref{Gain}.
An analytical form of $A_{11(22)}(\nu;\vect{k})$, obtained by Eqs.~(\ref{eq:GR}) and (\ref{eq:A}), enables us to conveniently study the renormalized band:
\begin{equation}
A_{11(22)}(\nu;\vect{k}) = 2|u_{\vect{k}}|^2\frac{\gamma}{(\nu-\Xt{eh}{k}^{-} \mp \En{k})^2+\gamma^2}
            +2|v_{\vect{k}}|^2\frac{\gamma}{(\nu-\Xt{eh}{k}^{-} \pm \En{k})^2+\gamma^2}.
\label{eq:A2}
\end{equation}
Here, $u_{\vect{k}}$ and $v_{\vect{k}}$ are the Bogoliubov coefficients defined as 
\begin{equation}
u_{\vect{k}} \equiv \sqrt{\frac{1}{2}+\frac{\Xt{eh}{k}^{+}}{2\En{k}}}, \hspace{0.5cm}
v_{\vect{k}} \equiv e^{\ii\theta_{\vect{k}}}\sqrt{\frac{1}{2}-\frac{\Xt{eh}{k}^{+}}{2\En{k}}},
\label{eq:Bogoliubov}
\end{equation}
with $\theta_{\vect{k}} \equiv \arg(\De{k})$.
These equations have remarkable similarities to the BCS theory in superconductors~\cite{Abrikosov75,Yamaguchi13}.
Therefore, it is clear that the gaps are opened around $\pm\mu/2$ with the magnitude of $\min[2\En{k}]$ when $\Xt{eh}{k}^{-}=0$ i.e. $\W{e}{k} = \W{h}{k}$ with $\mu^{\mathrm{B}}_{\mathrm{e}} = \mu^{\mathrm{B}}_{\mathrm{h}}$.
Note, however, that the unknown variables contained in Eqs.~(\ref{eq:A2}) and (\ref{eq:Bogoliubov}) are determined by the BEC-BCS-LASER crossover theory (Eqs.~(\ref{eq:MSBE_a0})-(\ref{eq:MSBE_nehk}) with Eqs.~(\ref{eq:RTA'})-(\ref{eq:A})) rather than the BCS theory.
In the BCS phase, the existence of the gap around $\pm\mu/2$ means the formation of Cooper pairs around the Fermi level because $\pm\mu/2$ is equivalent to the Fermi level.
In contrast, in the lasing phase, $\pm\mu/2$ corresponds to the laser frequency.\footnote{The origin of the gap is analogous to the Rabi splitting in resonance fluorescence~\cite{Scully97,Schmitt-Rink88,Henneberger92,Yamaguchi13}.}
Thus, the gap indicates the formation of bound e-h pairs by mediating photons around the laser frequency.
The semiconductor laser in Fig.~\ref{BEC-BCS-LASER} is drawn along this picture, where the e-h pairs are explicitly depicted.

Such a ``lasing gap'' is, at least in principle, measureable in the optical gain spectrum $G(\omega)$ by irradiating probe light with frequency $\omega$ because $G(\omega)$ is strongly affected by the renormalized band structure in general.
As a result, in the gain spectrum of the lasing phase (Fig.~\ref{Gain}(B)), there appears a transparent region originating from the gap.
The optical gain spectrum is thus one of important ways for the verification of the lasing gap.

In addition, we note that behaviors of the gain spectra vary drastically when the exciton-polariton BEC is changed to the laser phase.
By comparing Figs.~\ref{Gain}(A) and (B), for example, one can find only absorption but no gain in Fig.~\ref{Gain}(A).
This is mainly because there is no $\vect{k}$-region with inverted population $N_{k}>0$ $(\Leftrightarrow n_{\mathrm{h},k} = n_{\mathrm{e},k} > 0.5)$ in Fig.~\ref{Results}(A).
In contrast, optical gain is caused in Fig.~\ref{Gain}(B) because there are $\vect{k}$-regions with $N_{\vect{k}} > 0$ in Fig.~\ref{Results}(B).
Thus, the existence of the gain after the second threshold gives us important information to identify the phases in the system.\footnote{In fact, the second threshold and the blue-shift can also be caused by a different mechanism even if the system remains in quasi-equilibrium.
In this situation, however, the gain spectrum shows only absorption~\cite{Yamaguchi12}.
}

%%------------------------------------------------------------------------
%% Conclusions and perspective
%%------------------------------------------------------------------------
\section{Conclusions and perspectives}\label{sec:Conclusions}
In this contribution, we have presented a brief explanation of the BCS theory and the MSBE in the exciton-polariton system, to highlight their similarities and differences.
We have then shown a framework of describing the BCS theory (the BEC and BCS phases) and the MSBE (the semiconductor laser) in a unified way.
As a result, the existence of bound e-h pairs in the lasing phase as well as the lasing gap have been pointed out.
The results presented here are the physics elucidated for the first time by considering the BEC, BCS, and Laser phases in a unified way.
However, for example, effects of spontaneous emission~\cite{Scully97} and pure dephasing~\cite{Yamaguchi12-2} are still unclear.
In this respect, further studies are needed for a full understanding of this system.
Experimantal studies are also important, in particular, in a high density regime~\cite{Balili09,Nelsen09,Dang98,Tempel12-1,Tempel12-2,Tsotsis12,Kammann12,Horikiri13}.

Although we have focused on the exciton-polariton system in this contribution, we finally would like to emphasize that this system has a close relationship with superconductors and the Feshbach resonance in cold atom systems because interacting Fermi and Bose particles play important roles in the formation of ordered phases.
In this sense, it would be interesting to study the lasing gap by terahertz pulses in a manner similar to superconductors~\cite{Matsunaga13,Papenkort07}.
Inclusions of the e-h center-of-mass fluctuations with mass imbalance are also important, as discussed in the cold atom systems~\cite{Hanai13-1}, because these effects cannot be taken into account within the mean-field approximation.
We further note that fundamental problems of the nonequilibrium statistical physics are also included in this system in the sense of providing a bridge between the equilibrium and the nonequilibrium phases.
We hope that our approach also stimulates new studies in a wide range of such fields.

%%------------------------------------------------------------------------
%% Acknowledgement
%%------------------------------------------------------------------------
\begin{acknowledgement}
The authors are grateful to K. Kamide, R. Nii, Y. Yamamoto, T. Horikiri, Y. Shikano, Y. Matsuo, T. Yuge, and M. Bamba for fruitful discussions.
This work is supported by the JSPS through its FIRST Program, and DYCE, KAKENHI No. 20104008.
\end{acknowledgement}

%%------------------------------------------------------------------------
%% Appendix I: Excitonic effects in the low density limit
%%------------------------------------------------------------------------
\section*{Appendix I: Excitonic effects in the low density limit}
\addcontentsline{toc}{section}{Appendix}
In semiconductor exciton-polariton systems, the excitonic effects play quite important roles in the formation of Coulomb-bound e-h pairs (excitons) and exciton-polaritons.
In this Appendix, we, therefore, confirm the excitonic effects in our formalism~\footnote{Discussion in Appendix I is reproduced from the supplemental material in Ref.~\cite{Yamaguchi13}.}.

For this purpose, we now assume that the density of electrons and holes are sufficiently low ($n_{\mathrm{e},\vect{k}}, n_{\mathrm{h},\vect{k}} \ll 1$ or $N_{\vect{k}} \cong -1$) with no pumping and loss ($\gamma = 0$ and $\kappa = 0$).
Under this condition, $2\tilde\xi^{+}_{\mathrm{eh},\vect{k}}$ in Eq.~(\ref{eq:MSBE_pk}) can be written as
\begin{align}
2\tilde\xi^{+}_{\mathrm{eh},\vect{k}} &= \epsilon_{\mathrm{e},\vect{k}} + \epsilon_{\mathrm{h},\vect{k}} - \mu = \frac{k^2}{2m_{\mathrm{r}}}+E_g-\mu,
\label{eq:temp1}
\end{align}
where $1/m_{\mathrm{r}}=1/m_{\mathrm{e}}+1/m_{\mathrm{h}}$.
Then, Eqs.~(\ref{eq:MSBE_a0}) and (\ref{eq:MSBE_pk}) can be described as
\begin{align}
0&=-(\epsilon_{\mathrm{ph},0}-\mu)a_0+ g \sum_{\vect{k}} p_{\vect{k}},
\label{eq:a0_}\\
0&=-\left( \frac{k^2}{2m_{\mathrm{r}}}+E_g-\mu \right) p_{\vect{k}} + g^{*}a_0 + \sum_{\vect{k}'}U'_{\vect{k}'-\vect{k}}p_{\vect{k}'},
\label{eq:pk_}
\end{align}
where the definition of $\De{k} \equiv g^{*}a_0 + \sum_{\vect{k}'}U'_{\vect{k}-\vect{k}'}p_{\vect{k}'}$ is used.
Especially, for $g = 0$ in Eq.~\eqref{eq:pk_}, we obtain
\begin{align}
\frac{k^2}{2m_{\mathrm{r}}} p_{\vect{k}} - \sum_{\vect{k}'}U'_{\vect{k}'-\vect{k}}p_{\vect{k}'}=-(E_g-\mu) p_{\vect{k}},
\label{eq:Wannier}
\end{align}
which is nothing but the Schr$\ddot{\mathrm{o}}$dinger equation in $\vect{k}$-space for the single exciton bound state \cite{Yamaguchi12,Comte82,HaugKoch}.
This means that the Coulomb-bound e-h pairs (excitons) can be formed in the low density limit in the presented formalism.
In such a case, $p_{\vect{k}}$ can be described by the bound state e-h pair wave-function $\phi_{\vect{k}}$ ($p_{\vect{k}} = \eta \phi_{\vect{k}}$ with $\sum_{\vect{k}}|\phi_{\vect{k}}|^2 = 1$) with $\mu = E_{\mathrm{ex}}$, where $E_{\mathrm{ex}}$ is the energy level of the exciton and the binding energy corresponds to $E_{\mathrm{g}} - E_{\mathrm{ex}}$.
The formation of the exciton is, thus, includeed in the theory.

Although $p_{\vect{k}}$ is changed from the exciton wave-function $\phi_{\vect{k}}$ by the photon-mediated attraction in the case of $g \neq 0$, it is instructive to consider the case where such an effect is not so large.
In this limit, by substituting $p_{\vect{k}} \cong \eta\phi_{\vect{k}}$ into Eqs.~\eqref{eq:a0_} and \eqref{eq:pk_}, we obtain 
\begin{align}
0&=(\mu - E_{\mathrm{cav}})a_0 + g_{\mathrm{ex}}\eta, \\
0&=(\mu - E_{\mathrm{ex}})\eta + g_{\mathrm{ex}}^{*}a_0, 
\label{eq:eigen}
\end{align}
where $g_{\mathrm{ex}} \equiv g\sum_{\vect{k}}\phi_{\vect{k}}=g\phi_{\mathrm{ex}}(\vect{r}=0)$ is the coupling constant renormalized by the exciton wave-function.
Then, $\mu$ is given by one of the eigenvalues of these two coupled equations, which are the eigen-energies of the upper and lower polaritons:
\begin{align}
E_{\mathrm{UP/LP}} = \frac{E_{\mathrm{cav}}+E_{\mathrm{ex}} \pm \sqrt{(E_{\mathrm{cav}}-E_{\mathrm{ex}})^2+4|g_{\mathrm{ex}}|^2}}{2}. 
\label{eq:UPLP}
\end{align}
Here, $E_{\mathrm{UP}}$ and $E_{\mathrm{LP}}$ in Eq.~\eqref{eq:UPLP} are the well-known expressions obtained when the excitons are treated as simple bosons \cite{HaugKoch}.
This means that the formation of exciton-polaritons are also included in the theory.
The excitonic effects are, thus, taken into account in our formalism within the mean-field approximation.
We note that the procedure shown here is basically the same as Section 2.1.2 in Ref.~\cite{Yamaguchi12}.

%%------------------------------------------------------------------------
%% Appendix II: Proof of equivalence
%%------------------------------------------------------------------------
\section*{Appendix I\hspace{-.1em}I: Proof of equivalence}
\addcontentsline{toc}{section}{Appendix}
In the main text, we have mentioned that the formalism in Ref.~\cite{Yamaguchi12} seems quite different from the MSBE.
This formalism can be described by the follwoing simultaneous equations with the unknown variables of $\De{k}$, $\nek$, $\nhk$, and $\mu$:
\begin{multline}
\De{k} = \sum_{\vect{k}'} U^{\mathrm{eff},\kappa}_{\vect{k}',\vect{k}} \Ded{k} \int^{\infty}_{-\infty} \frac{\dd\nu}{2\pi} L_{\vect{k}'}(\nu) \\
		  \times \left \{ (F_{\mathrm{e}}^{\mathrm{B}}(\nu) + F_{\mathrm{h}}^{\mathrm{B}}(\nu)) (\nu-\tilde\xi^{-}_{\mathrm{eh},\vect{k}'}) 
		  + (F_{\mathrm{e}}^{\mathrm{B}}(\nu) - F_{\mathrm{h}}^{\mathrm{B}}(\nu)) (\tilde\xi^{+}_{\mathrm{eh},\vect{k}'} + \ii\gamma) \right\},
\label{eqa:Delta}
\end{multline}
\begin{equation}
n_{\mathrm{e(h)},\vect{k}} = \frac{1}{2} \mp \int^{\infty}_{-\infty} \frac{\dd\nu}{2\pi} L_{\vect{k}}(\nu)
		\left\{ F_{\mathrm{e(h)}}^{\mathrm{B}}(\nu)[(\nu\pm\tilde\xi_{\mathrm{h(e)},\vect{k}})^2+\gamma^2] + F_{\mathrm{h(e)}}^{\mathrm{B}}(\nu)|\De{k}|^2 \right\},
\label{eqa:nehk}
\end{equation}
where $U_{\vect{k}',\vect{k}}^{\mathrm{eff},\kappa} \equiv |g|^2/(\xi_{\mathrm{ph},0}-\ii\kappa)+U'_{\vect{k}'-\vect{k}}$ and
\begin{equation}
L_{\vect{k}}(\nu) \equiv \frac{\gamma}{[(\nu-\tilde\xi^{-}_{\mathrm{eh},\vect{k}}-\En{k})^{2}+\gamma^2][(\nu-\tilde\xi^{-}_{\mathrm{eh},\vect{k}}+\En{k})^{2}+\gamma^2]}.
\label{eqa:L}
\end{equation}
$F_{\mathrm{e}}^{\mathrm{B}}(\nu)$ and $F_{\mathrm{h}}^{\mathrm{B}}(\nu)$ are respectively defined as
\begin{eqnarray}
F_{\mathrm{e}}^{\mathrm{B}}(\nu) & \equiv & \tanh \left( \frac{\beta[\nu - \mu^{\mathrm{B}}_{\mathrm{e}} + \mu/2]}{2} \right) = 1-2f^{\mathrm{B}}_{\mathrm{e}}(\nu),\\
\label{eqa:Fe}
F_{\mathrm{h}}^{\mathrm{B}}(\nu) & \equiv & \tanh \left( \frac{\beta[\nu + \mu^{\mathrm{B}}_{\mathrm{h}} - \mu/2]}{2} \right) = 2f^{\mathrm{B}}_{\mathrm{h}}(-\nu)-1.
\label{eqa:Fh}
\end{eqnarray}
In this Appendix I\hspace{-.1em}I, therefore, we prove that Eqs.~(\ref{eqa:Delta})-(\ref{eqa:Fh}) are equivalent to Eqs.~(\ref{eq:MSBE_a0})-(\ref{eq:MSBE_nehk}) with Eqs.~(\ref{eq:RTA'})-(\ref{eq:A}) under the steady-state condition $\partial_t\EX{\hat{O}} = 0$. 

%------------------------------------------------------------------------------
For this purpose, we here note that the following sum rule is satisfied for the single-partice spectral function:
\begin{equation}
\int^{\infty}_{-\infty}\frac{\dd\nu}{2\pi}A_{\alpha\alpha'}(\nu;\vect{k}) = \delta_{\alpha,\alpha'}.
\label{eqa:SumRule}
\end{equation}
This relation can be confirmed by the direct integration of $A_{\alpha\alpha'}(\nu;\vect{k})$ described by elements of a matrix 
\begin{align}
A(\nu;\vect{k}) &= \frac{-2}{|D_{\vect{k}}(\nu)|^2}\left(
\begin{array}{cc}
\operatorname{Im}[D^{*}_{\vect{k}}(\nu)(\nu+\tilde\xi_{\mathrm{h},\vect{k}}+\ii\gamma)] & \operatorname{Im}[D_{\vect{k}}(\nu)]\De{k} \\
\operatorname{Im}[D_{\vect{k}}(\nu)]\De{k}^{*} & \operatorname{Im}[D^{*}_{\vect{k}}(\nu)(\nu-\tilde\xi_{\mathrm{e},\vect{k}}+\ii\gamma)] \\
\end{array}\right) \nonumber \\
&= \frac{-2\gamma}{|D_{\vect{k}}(\nu)|^2}\left(
\begin{array}{cc}
-[(\nu+\tilde\xi_{\mathrm{h},\vect{k}})^2+\gamma^2+|\De{k}|^2] & 2\De{k}(\nu-\tilde\xi^{-}_{\mathrm{eh},\vect{k}}) \\
2\De{k}^{*}(\nu-\tilde\xi^{-}_{\mathrm{eh},\vect{k}}) & -[(\nu - \tilde\xi_{\mathrm{e},\vect{k}})^2+\gamma^2+|\De{k}|^2] \\
\end{array}\right),
\label{eqa:A}
\end{align}
which is obtained from the definition of Eq.~(\ref{eq:A}) with Eq.~(\ref{eq:GR}):
\begin{equation}
G^{\mathrm{R}}_{\vect{k}}(\nu) = \frac{1}{|D_{\vect{k}}(\nu)|^2}\left(
\begin{array}{cc}
D^{*}_{\vect{k}}(\nu)(\nu+\tilde\xi_{\mathrm{h},\vect{k}}+\ii\gamma) & -D^{*}_{\vect{k}}(\nu)\De{k} \\
-D^{*}_{\vect{k}}(\nu)\De{k}^{*} & D^{*}_{\vect{k}}(\nu)(\nu-\tilde\xi_{\mathrm{e},\vect{k}}+\ii\gamma) \\
\end{array}\right),
\label{eqa:GR}
\end{equation}
where 
\begin{equation}
D_{\vect{k}}(\nu) \equiv (\nu-\tilde\xi^{-}_{\mathrm{eh},\vect{k}} + \En{k} + \ii\gamma)(\nu-\tilde\xi^{-}_{\mathrm{eh},\vect{k}} - \En{k} + \ii\gamma).
\end{equation}
The diagonal element $A_{11(22)}(\nu;\vect{k})$ can then be described as Eq.~(\ref{eq:A2}).
In the following, by using thse expressions, Eqs.~(\ref{eqa:Delta}) and (\ref{eqa:nehk}) are derived from Eqs.~(\ref{eq:MSBE_a0})-(\ref{eq:MSBE_nehk}) with Eqs.~(\ref{eq:RTA'})-(\ref{eq:A}).

%----------------- nu-integral forms of Nk and pk -----------------------
\runinhead{$\nu$-integral forms of $N_{\vect{k}}$ and $\pk$:}
First, we discuss $\nu$-integral forms of the population inversion $N_{\vect{k}} \equiv \nek + \nhk -1$ and the polarization function $\pk$ because Eqs.~(\ref{eqa:Delta}) and (\ref{eqa:nehk}) are described by the integration with respect to $\nu$.
From Eqs.~(\ref{eq:MSBE_nehk}) and (\ref{eq:RTA'}) with $\partial_t n_{\mathrm{e(h)},\vect{k}}=0$, we obtain
\begin{equation}
N_{\vect{k}} = -\frac{2}{\gamma}\operatorname{Im}[\De{k}\pk^{*}]
			+ \int^{\infty}_{-\infty}\frac{\dd\nu}{2\pi} \left\{ f^{\mathrm{B}}_{\mathrm{e}}(\nu)A_{11}(\nu;\vect{k})
			-(1-f^{\mathrm{B}}_{\mathrm{h}}(-\nu))A_{22}(\nu;\vect{k})   \right\}.
\label{eqa:PopInv1}
\end{equation}
where Eq.~(\ref{eqa:SumRule}) is used.
In a similar manner, from Eqs.~(\ref{eq:MSBE_pk}) and (\ref{eq:RTA'}) with $\partial_t \pk=0$,
\begin{align}
\pk = &- \frac{\De{k}}{2(\tilde\xi^{+}_{\mathrm{eh},\vect{k}} - \ii\gamma)}N_{\vect{k}} \nonumber \\
	  &+ \frac{\gamma}{\tilde\xi^{+}_{\mathrm{eh},\vect{k}} - \ii\gamma} 
		\int^{\infty}_{-\infty}\frac{\dd\nu}{2\pi}\left\{ [1-f^{\mathrm{B}}_{\mathrm{h}}(-\nu)]G^{\mathrm{R}}_{12}(\nu;\vect{k})
		-f^{\mathrm{B}}_{\mathrm{e}}(\nu)G^{\mathrm{R}*}_{21}(\nu;\vect{k}) \right\}.
\label{eqa:pk1}
\end{align}
Therefore, with Eqs.~(\ref{eqa:A}) and (\ref{eqa:GR}), substitution of Eq.~(\ref{eqa:pk1}) into Eq.~(\ref{eqa:PopInv1}) yields
\begin{multline}
N_{\vect{k}}=\int^{\infty}_{-\infty}\frac{\dd\nu}{2\pi}\frac{2\gamma}{|D_{\vect{k}}(\nu)|^2}
			\left \{ f^{\mathrm{B}}_{\mathrm{e}}(\nu) [(\nu+\tilde\xi_{\mathrm{h},\vect{k}})^2-|\De{k}|^2+\gamma^2] \right. \\
			\left. +[f^{\mathrm{B}}_{\mathrm{h}}(-\nu)-1] [(\nu-\tilde\xi_{\mathrm{e},\vect{k}})^2-|\De{k}|^2+\gamma^2] \right \}.
\label{eqa:PopInv2}
\end{multline}
By substituting Eq.~(\ref{eqa:PopInv2}) into Eq.~(\ref{eqa:pk1}), we also find
\begin{equation}
\pk=\De{k} \int^{\infty}_{-\infty}\frac{\dd\nu}{2\pi} \frac{2\gamma}{|D_{\vect{k}}(\nu)|^2}
			\left \{ [f^{\mathrm{B}}_{\mathrm{h}}(-\nu)-1](\nu - \tilde\xi_{\mathrm{e},\vect{k}} - \ii\gamma)
			- f^{\mathrm{B}}_{\mathrm{e}}(\nu)(\nu + \tilde\xi_{\mathrm{h},\vect{k}} + \ii\gamma)   \right \}.
\label{eqa:pk2}
\end{equation}
Although the derivation of Eqs.~(\ref{eqa:PopInv2}) and (\ref{eqa:pk2}) is straightforward, the following equations would be useful in the derivation:
\begin{align}
\pm|\De{k}|^2\operatorname{Im}[D_{\vect{k}}(\nu)(\tilde\xi^{+}_{\mathrm{eh},\vect{k}}-\ii\gamma)]
 + &([\tilde\xi^{+}_{\mathrm{eh},\vect{k}}]^2+\gamma^2)\operatorname{Im}[D^{*}_{\vect{k}}(\nu)(\nu \mp \tilde\xi_{\mathrm{e(h)},\vect{k}}+\ii\gamma)] \nonumber \\
 &= -\gamma(\En{k}^2 + \gamma^2)((\nu \mp \tilde\xi_{\mathrm{e(h)},\vect{k}})^2-|\De{k}|^2+\gamma^2), \\
(\nu+\tilde\xi_{\mathrm{h}})^2 - |\De{k}|^2 + \gamma^2 &-D_{\vect{k}}(\nu) = 2 (\tilde\xi^{+}_{\mathrm{eh},\vect{k}} - \ii\gamma)(\nu + \tilde\xi_{\mathrm{h}} + \ii\gamma), \\
(\nu-\tilde\xi_{\mathrm{e}})^2 - |\De{k}|^2 + \gamma^2 &-D^{*}_{\vect{k}}(\nu) = -2 (\tilde\xi^{+}_{\mathrm{eh},\vect{k}} - \ii\gamma)(\nu - \tilde\xi_{\mathrm{e}} - \ii\gamma). 
\end{align}
The $\nu$-integral forms of $N_{\vect{k}}$ and $\pk$ are thus obtained as Eqs.~(\ref{eqa:PopInv2}) and (\ref{eqa:pk2}), respectively.
These expressions are helpful to find the $\nu$-integral forms of $\De{k}$, $\nek$, and $\nhk$, which turn out to be the same as Eqs.~(\ref{eqa:Delta})-(\ref{eqa:Fh}), as shown below.

%----------------- Derivation of Delta_k -----------------------
\runinhead{Derivation of $\De{k}$:}
From the definition of $\De{k} \equiv g^{*}a_0 + \sum_{\vect{k}'}U'_{\vect{k}-\vect{k}'}p_{\vect{k}'}$ and Eq.~(\ref{eq:MSBE_a0}) with $\partial_t\a0 = 0$, $\De{k}$ can be described as 
\begin{equation}
\De{k} = \sum_{\vect{k}'} \left\{ \frac{|g|^2}{\xi_{\mathrm{ph},0}-\ii\kappa}+U'_{\vect{k}'-\vect{k}} \right\}p_{\vect{k}'}
	   = \sum_{\vect{k}'} U_{\vect{k}',\vect{k}}^{\mathrm{eff},\kappa} p_{\vect{k}'}.
\label{eqa:temp2}
\end{equation}
Therefore, after the substitution of Eq.~(\ref{eqa:pk2}) into Eq.~(\ref{eqa:temp2}), we obtain
\begin{multline}
\De{k} = \sum_{\vect{k}'} U_{\vect{k}',\vect{k}}^{\mathrm{eff},\kappa} \Delta_{\vect{k}'} \int^{\infty}_{-\infty}\frac{\dd\nu}{2\pi} L_{\vect{k}'}(\nu)
	\left \{ [F^{\mathrm{B}}_{\mathrm{h}}(\nu)-1][\nu - \tilde\xi_{\mathrm{e},\vect{k}'} - \ii\gamma] \right. \\
	\left. + [F^{\mathrm{B}}_{\mathrm{e}}(\nu)-1][\nu + \tilde\xi_{\mathrm{h},\vect{k}'} + \ii\gamma] \right\},
\end{multline}
where the definitions of Eqs.~(\ref{eqa:L})-(\ref{eqa:Fh}) are used with $L_{\vect{k}}(\nu) = \gamma/|D_{\vect{k}}(\nu)|^2$.
This equation can be rewritten as
\begin{multline}
\De{k} = \sum_{\vect{k}'} U^{\mathrm{eff},\kappa}_{\vect{k}',\vect{k}} \Ded{k} \int^{\infty}_{-\infty} \frac{\dd\nu}{2\pi} L_{\vect{k}'}(\nu) 
		    \left \{ (F_{\mathrm{e}}^{\mathrm{B}}(\nu) + F_{\mathrm{h}}^{\mathrm{B}}(\nu) - 2) (\nu-\tilde\xi^{-}_{\mathrm{eh},\vect{k}'}) \right. \\
		  + \left. (F_{\mathrm{e}}^{\mathrm{B}}(\nu) - F_{\mathrm{h}}^{\mathrm{B}}(\nu)) (\tilde\xi^{+}_{\mathrm{eh},\vect{k}'} + \ii\gamma) \right\}.
\label{eqa:Delta2}
\end{multline}
By noting $\int \frac{\dd\nu}{2\pi} L_{\vect{k}'}(\nu) (\nu-\tilde\xi^{-}_{\mathrm{eh},\vect{k}'}) = 0$, we thus find that Eq.~(\ref{eqa:Delta2}) is equivalent to Eq.~(\ref{eqa:Delta}).

%----------------- Derivation of nek and nhk -----------------------
\runinhead{Derivation of $\nek$ and $\nhk$:}
Our remaining task is now to derive the $\nu$-integral forms of $\nek$ and $\nhk$.
By multiplying $\De{k}$ by the complex conjugate of Eq.~(\ref{eqa:pk2}),
\begin{equation}
\frac{1}{\gamma}\operatorname{Im}[\De{k}\pk^{*}] = |\De{k}|^2 \int^{\infty}_{-\infty}\frac{\dd\nu}{2\pi} \frac{2\gamma}{|D_{\vect{k}}(\nu)|^2} \{ f^{\mathrm{B}}_{\mathrm{e}}(\nu) + f^{\mathrm{B}}_{\mathrm{h}}(-\nu) -1 \},
\label{eqa:temp1}
\end{equation}
can be obtained.
The substitution of Eq.~(\ref{eqa:temp1}) into Eq.~(\ref{eq:MSBE_nehk}) with $\partial_t n_{\mathrm{e(h)},\vect{k}} = 0$, then, yields 
\begin{multline}
n_{\mathrm{e(h)},\vect{k}} = \int^{\infty}_{-\infty} \frac{\dd\nu}{2\pi} \frac{2\gamma}{|D_{\vect{k}}(\nu)|^2}
	\left \{ [(\nu \pm \tilde\xi_{\mathrm{h(e)},\vect{k}})^2 + \gamma^2]f^{\mathrm{B}}_{\mathrm{e(h)}}(\pm\nu) \right. \\
	\left. -|\De{k}|^2[ f^{\mathrm{B}}_{\mathrm{h(e)}}(\mp\nu) -1] \right\},
\end{multline}
which can be rewritten as 
\begin{multline}
n_{\mathrm{e(h)},\vect{k}} = \int^{\infty}_{-\infty} \frac{\dd\nu}{2\pi} L_{\vect{k}}(\nu)
	\left \{ [(\nu \pm \tilde\xi_{\mathrm{h(e)},\vect{k}})^2 + \gamma^2 + |\De{k}|^2] \right. \\
	\left. \mp [(\nu \pm \tilde\xi_{\mathrm{h(e)},\vect{k}})^2 + \gamma^2]F^{\mathrm{B}}_{\mathrm{e(h)}}(\nu) \mp |\De{k}|^2F^{\mathrm{B}}_{\mathrm{h(e)}}(\nu) \right\}.
\label{eqa:nehk2}
\end{multline}
We then find that Eq.~(\ref{eqa:nehk2}) is identical to Eq.~(\ref{eqa:nehk}) because, from Eqs.~(\ref{eqa:SumRule}) and (\ref{eqa:A}), 
\begin{equation}
\int^{\infty}_{-\infty} \frac{\dd\nu}{2\pi} L_{\vect{k}}(\nu)[(\nu \pm \tilde\xi_{\mathrm{h(e)},\vect{k}})^2 + \gamma^2 + |\De{k}|^2] = \frac{1}{2}.
\end{equation}
Thus, we have shown that Eqs.~(\ref{eqa:Delta}) and (\ref{eqa:nehk}) are derived from Eqs.~(\ref{eq:MSBE_a0})-(\ref{eq:MSBE_nehk}) with Eqs.~(\ref{eq:RTA'})-(\ref{eq:A}).
This means that the formalism in Ref.~\cite{Yamaguchi12} is equivalent to Eqs.~(\ref{eq:MSBE_a0})-(\ref{eq:MSBE_nehk}) with Eqs.~(\ref{eq:RTA'})-(\ref{eq:A}).

%%------------------------------------------------------------------------
%% Reference
%%------------------------------------------------------------------------
% \bibliography{ref-yamaguchi.bib}

\end{document}